\documentclass[twocolumn,showpacs,showkeys,preprintnumbers,
               amsmath,amssymb,floatfix,10pt]{revtex4}
\usepackage{graphicx}
\usepackage{txfonts}
\usepackage[small]{titlesec}
\titlespacing*{\section}{0pt}{0.1cm}{0.1cm}
\begin{document} 
   \title{A singular mutation in the hemagglutinin of the 1918 pandemic virus}

   \author{Yves-Henri Sanejouand}

   \affiliation{UFIP, CNRS UMR 6286,\\Facult\'e des Sciences et des Techniques,\\Nantes, France.\\
              \email{yves-henri.sanejouand@univ-nantes.fr}}
   \date{October 14, 2016}
   \maketitle

\section*{Abstract}

The influenza pandemic of 1918-1919 killed at least 50 million people.
The reasons why this pandemic was so deadly remain
largely unknown~\cite{Mueller:02}. However, 
It has been shown that the 1918 viral hemagglutinin allows to reproduce the hallmarks of the
illness observed during the original pandemic~\cite{Kawaoka:04}.
Thanks to the wealth of hemagglutinin sequences accumulated over the last decades,
amino-acid substitutions that are found in the 1918-1919 sequences but rare otherwise can be identified with high confidence.
Such an analysis reveals that Gly188, which is located within a key motif of the receptor binding site,
is so rarely found in hemagglutinin sequences that, taken alone, it is likely to be deleterious.
Monitoring this singular mutation in viral sequences may help prevent another dramatic pandemic. 
 
\section*{Introduction}


A dozen years ago, it was shown that introducing the 1918 hemagglutinin (HA) confers enhanced pathogenicity in mice to recent human viruses that are otherwise non-pathogenic in this host.
Moreover, like the 1918 one, these recombinant viruses infect the entire lung and induce high levels of macrophage-derived chemokines and cytokines, which results in infiltration of inflammatory cells and severe haemorrhage~\cite{Kawaoka:04}.
In macaques, the whole 1918 virus causes a highly pathogenic respiratory infection that
culminates in acute respiratory distress and a fatal outcome~\cite{Kawaoka:07}.
Although the 1918 polymerase genes were found essential for optimal virulence, replacing the 1918 HA by a contemporary human one proved enough for abolishing the lethal outcome of the 1918 virus infection in mices~\cite{Garciasastre:05}, further underlining the key role of the 1918 HA in the deadly process.  

The first 1918 HA sequences were obtained in 1999 from formalin-fixed, 
paraffin-embedded lung tissue samples prepared during
the autopsy of victims of the influenza pandemic, as well as from a frozen
sample obtained by \textit{in situ} biopsy of the lung of a victim buried
in permafrost since 1918~\cite{Taubenberger:99}.

Since then, the number of HA sequences determined each year
has grown dramatically, jumping from $\approx$100 in the     
nineties to $\approx$3,000 per year nowadays~\footnote{According 
to the NCBI influenza virus resource.}.
The goal of the present work is to take advantage of this wealth of data
for identifying features that are unique to 
the 1918 sequence, the underlying hypothesis being that they
may prove responsible for the unique behaviour of viruses displaying the 1918 HA. 

\begin{figure}
\centering
\includegraphics[width=9.0 cm]{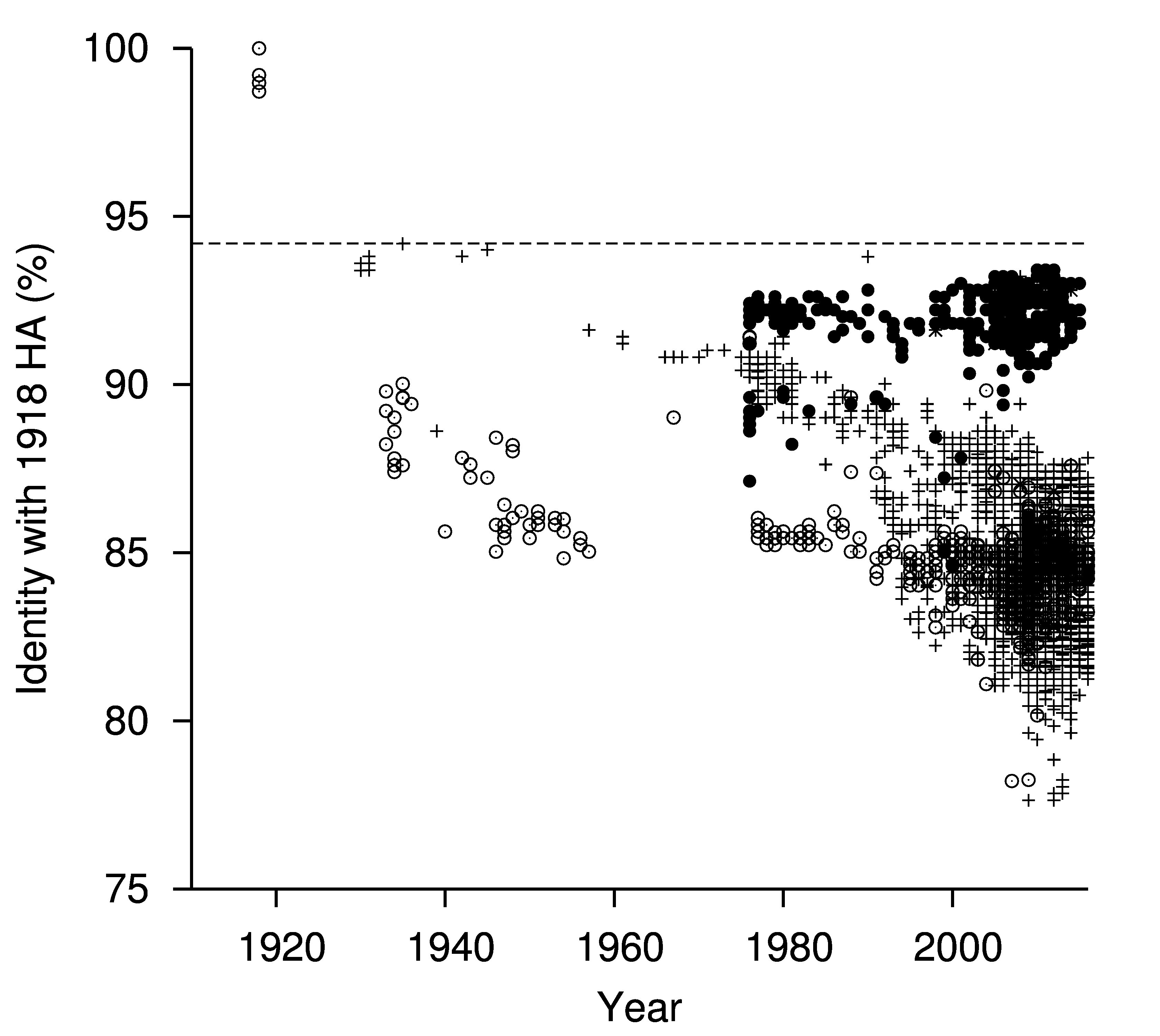}
\vskip -0.3 cm
\caption{\textit{Percentage of identity of influenza A hemagglutinin sequences with the 1918 sequence, as a function of time.} Viruses come from human (open circles), swine (plus symbols), avian (filled circles) or other hosts (stars). 
The dotted line indicates the highest level of sequence identity observed since 1918 (94.2\%).
Results are only shown for the H1 subtype, since other subtypes have sequence identities with the 1918 sequence that are below 75\%.}
\label{Fig:idoft}
\end{figure}

\section*{Methods}

52006 HA different protein sequences were retrieved~\footnote
{On September 6$^{th}$, 2016.}
from the NCBI influenza virus resource~\cite{Ivr:08}, 
sequences coming from laboratory viral strains being disregarded. 
Multiple pairwise sequence alignment was performed using BLAST~\cite{Blast:97}, 
version 2.2.19, taking as a reference the long H1 sequence from virus A/Thailand/CU-MV10/2010. MVIEW~\cite{Mview:98}, version 1.60.1, was used for converting the BLAST output into a standard multiple sequence alignement (MSA). In this MSA, 
3365 different amino-acid residues are observed at least 20 times,
that is, 6.2 per position along the sequence, on average~\footnote{Positions
with a gap in the majority of the sequences were excluded from this statistics.}. 
Since 1999, a dozen 1918 HA sequences have been determined,
at least partially.
They differ, at most, by a couple of mutations~\cite{Taubenberger:99,Taubenberger:03}.
One of them was chosen as a representative, namely, 
the sequence found in PDB structure 4EEF 
(with H3 subtype residue numbering)~\cite{Baker:12}.
 
\section*{Results}

As shown in Figure~\ref{Fig:idoft}, 
known HA sequences~\footnote{37268 sequences were considered for sequence identity comparisons, those with less than 400 residues being disregarded.} 
of post-1919 viruses 
are \textit{all} less than 95\% identical to the 1918 one, that is,
complete ones differ from the 1918 sequence by at least 25 amino-acid 
substitutions~\footnote{The HA sequence in PDB structure 4EEF is 498 residues long.}.
Strinkingly, after 1919,   
HA sequences of human viruses are less than    
91.4\% identical to the 1918 HA, a $\approx$90\% level of sequence identity being observed
in the thirties as well as more recently, though in a few instances only~\footnote{Four
cases in 1976, for instance, the other 13 human virus sequences determined that year 
being of another subtype (H3).}.
On the other hand, HA sequences of avian viruses more than 93\% identical to the 1918 HA
have been observed each year since 2005, suggesting that a selection pressure
favoring 1918 HA-like sequences is at work in avian species, in line with the hypothesis of an avian origin for the 1918-1919 pandemic~\cite{Taubenberger:99}. Moreover,
among the 41 post-1950 sequences that are more than 93\% identical to the 1918 HA,
35 (85\%) come from duck species~\footnote{The six remaining ones come from two other avian species, goose and sparrow, from swine -three cases, the last one being found in the environment.}, further pinpointing aquatic birds as a possible reservoir~\cite{Webster:96,Webster:15}. 

\begin{table}
\centering        
\begin{tabular}{c c c c c c}   
\hline\hline                
1918 HA     & Human & Avian & Swine  &   All &   All \\    
residue$^a$ &   H1  &   H1  &   H1   &    H1 &    HA \\
\hline                        
Gly 188$^b$ &   0   &    0  &   47   &    47 &    82 \\
Arg \textbf{116}$^c$ &   4   &  438  &  162   &   621 &  2071 \\   
Ser 159$^b$ &  17   &   21  &   47   &    85 &  4066 \\
His 285$^b$ &  17   &  433  &  802   &  1269 &  1297 \\
Ala 219$^b$ &  22   &  458  & 1642   &  2141 &  5964 \\
Asn 72$^c$  &  22   &  468  &  289   &   796 &  7489 \\
Lys \textbf{120}$^b$ &  29   &  446  &  164   &   656 & 10549 \\
Arg \textbf{153}$^c$ &  31   &   39  &  418   &   488 & 22855 \\
Ser 264$^b$ &  45   &  463  & 3562   &  4090 &  9738 \\
Leu 70$^b$  &  48   &  439  & 2857   &  3367 & 25266 \\
\hline
Asp 190$^b$ & 10055 &   24  & 5340   & 15436 & 28495 \\
Ser 193$^b$ &  7238 &  312  & 2758   & 10334 & 15936 \\
Lys \textbf{222}$^b$ & 10871 &  432  & 6101   & 17437 & 23284 \\
Asp 225$^b$ &  9940 &   12  & 4183   & 14150 & 18093 \\
\hline
Total       & 11062 &  508  & 6183   & 17794 & 51992 \\
\hline                                   
\end{tabular}
\caption{\textit{Number of post-1919 hemagglutinin sequences with same residue as the 
1918 sequence.} Top: residues found in less than 50 human H1 sequences. 
Bottom: key residues involved in receptor binding~\cite{Skehel:04}. 
Bold: residue index of an highly conserved residue, that is,
a residue found in more than 95\% of the H1 sequences.
$^a$H3 subtype residue numbering; $^b$HA1 subunit; $^c$HA2 subunit. 
}            
\label{tbl:mutations}   
\end{table}

However, the 1918 HA sequence is a singular one.
For instance, 17 amino-acid residues in this sequence are found in less
than 1\% of other human H1 sequences. The ten less frequent ones are shown on top of Table~\ref{tbl:mutations}. Most of them are frequent in avian H1 sequences,
or often found~\footnote{1297 cases at least.} in sequences of other HA subtypes (last column). 

As a striking exception, after 1919, Gly 188
has \textit{not} been observed again in human H1 sequences.
It has also \textit{not} been observed in avian ones.
As a matter of fact, it has only been observed in 47 H1 sequences, all of them from swine,
a single time in 2003, once each year between 2009 and 2012, and
several times each year since then~\footnote{11 times in 2014, 14 times in 2015.}.  
In human HA of other subtypes, it has been observed 11 times, in sequences from H3N2 or H5N1 viruses. 
Since 1919, Gly 188 has \textit{only} been observed 82 times, the first time in 2000, in the sequence of an avian H9 HA. 


Residue 188 is located at the N-terminus of the "190-helix", which is involved in the HA receptor binding site, but it
does not interact directly with the receptor~\cite{Skehel:04,Skehel:00}.
Three residues are usually observed at this position, 
namely, serine (41\%), isoleucine (33\%) and threonine (21\%).
Interestingly, proline which, like glycine, can have a direct impact
on the secondary structure~\cite{Balaram:98},
the folding~\cite{Kiefhaber:05} or the stability~\cite{Matthews:92} of a protein, 
is also rarely observed~\footnote{10 times in all HA sequences.}.
This suggests that, taken alone, Gly 188 is likely to be deleterious.
However, since the 1918 virus proved efficient,
one or several compensatory mutations have to be present in its HA sequence.

\begin{figure}
\centering
\includegraphics[width=8.5 cm]{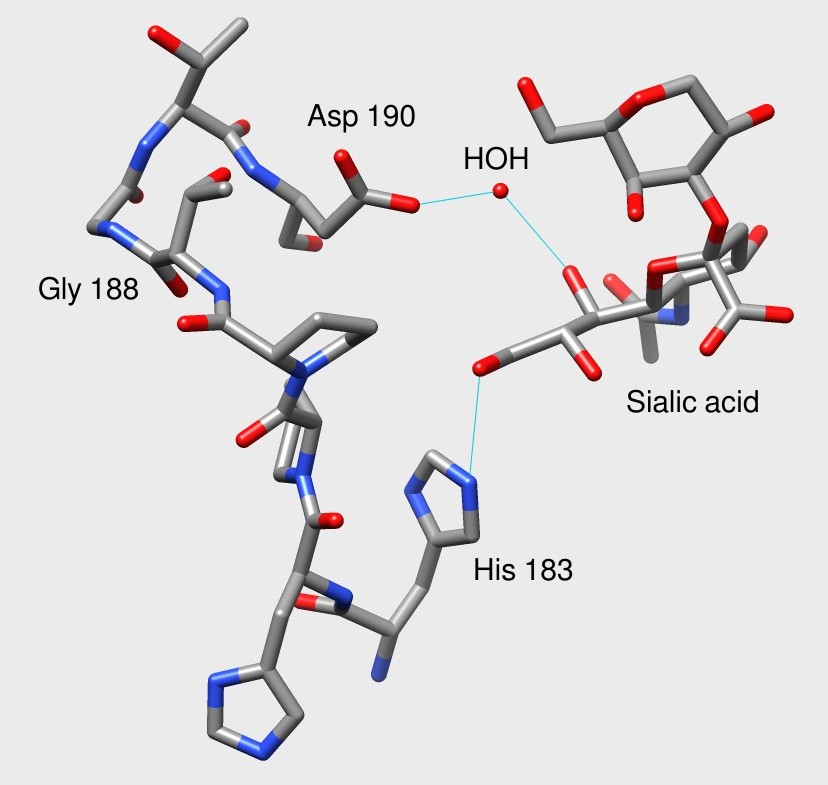}
\vskip -0.3 cm
\caption{\textit{The 188-hook}. Gly 188 is located between two residues interacting either directly (His 183) or through a water molecule (Asp 190) with the sialic acid moiety of the hemagglutinin receptor. The conformation shown comes from PDB structure 4JUH~\cite{Gao:13}. Drawn with UCSF Chimera~\citep{Chimera}.}
\label{Fig:hook}
\end{figure}

\section*{Discussion}

\textbf{Why has this mutation been overlooked ?}
After 1919, Gly 188 has \textit{not} been found again in human H1 sequences.
Overall, it has been found in \textit{only} 0.2\% of all known HA sequences.
The reason why this singular mutation seems to have been overlooked
is probably the following one: in the eleven 1918-1919 HA sequences known so far,
Gly 188 has been found ten times. In other words, there is an exception, namely, the
HA sequence of virus A/London/1/1918~\cite{Taubenberger:03}.

\noindent
\textbf{How may it be involved in the deadly process ?}
Gly 188 is located between His 183 and Asp 190, two key residues of the receptor binding site (Figure~\ref{Fig:hook}). His 183 is highly conserved, being found in 99.9\% of H1 sequences. It is thus likely to be involved in the specific recognition of the sialic acid moiety of the HA receptor.    
On the other hand, Asp 190 is well conserved in H1 HA (88\%) but much less in all HA (56\%), being frequently replaced by Glu (32\%). 
This suggests that it is involved in more subtle aspects of the recognition. Indeed, a D190E mutation in the 1918 HA results in a preference for the $\alpha2,3$ sialic acid (avian) receptor~\cite{Palese:05}. 
Thus, given the inherent flexibility of Gly residues, introducing Gly
at position 188 may allow for alternative orientations of Asp 190
and, as a consequence, for the recognition of other conformations of the sialic acid moiety,
or to sialic acid moities linked to di-saccharides other than Gal-GlcNac.
Though a glycan microarray analysis
confirmed a specific recognition of $\alpha2,6$ sialic acid receptors by 1918
HA, with no binding when GlcNac is absent~\cite{Wilson:06},
a change in its repertoire of HA receptors could indeed explain why the 1918 virus proved so deadly.  

\section*{Conclusion}

Gly 188 has rarely been observed in HA sequences (Table~\ref{tbl:mutations}). Since it is located within a key motif of the HA receptor binding site (Figure~\ref{Fig:hook}), this mutation may have an impact on receptor recognition and specificity. This, in turn, could explain why the 1918 HA confers enhanced pathogenicity.  
It is thus important to check the effects of this mutation on receptor function. 
Meanwhile, it may prove important to monitor this mutation, noteworthy in swine viruses since it has been observed in swine H1 sequences each year since 2009.


\end{document}